\documentclass[twocolumn,prl,superscriptaddress,showpacs,amsmath,amssymb,aps,10pt]{revtex4-1}

\usepackage{graphicx}
\usepackage{color}
\newcommand{\mosto}{MoS$_{\mathrm{2}}$}

\begin{document}

\title{Field effect in stacked van der Waals heterostructures: Stacking sequence matters}

\author{Daniele Stradi}
\email[]{daniele.stradi@quantumwise.com}
\affiliation{Quantumwise A/S, Fruebjergvej 3, Postbox 4, DK-2100 Copenhagen, Denmark}
\affiliation{Center for Nanostructured Graphene (CNG), Department of Micro- and Nanotechnology (DTU Nanotech), Technical University of Denmark, DK-2800, Kgs. Lyngby, Denmark}

\author{Nick R. Papior}
\affiliation{ICN2 - Institut Catal\`a de Nanoci\`encia i Nanotecnologia, Campus UAB, 08193 Bellaterra, Spain}
\affiliation{Center for Nanostructured Graphene (CNG), Department of Micro- and Nanotechnology (DTU Nanotech), Technical University of Denmark, DK-2800, Kgs. Lyngby, Denmark}

\author{Mads Brandbyge}
\affiliation{Center for Nanostructured Graphene (CNG), Department of Micro- and Nanotechnology (DTU Nanotech), Technical University of Denmark, DK-2800, Kgs. Lyngby, Denmark}

\date{\today}

\begin{abstract}
Stacked van der Waals (vdW) heterostructures where semi-conducting two-dimensional (2D) materials are contacted by overlayed graphene electrodes enable atomically-thin, flexible electronics.
We use first-principles quantum transport simulations of graphene-contacted MoS$_\mathrm{2}$ devices to show how the transistor effect critically depends on the stacking configuration relative to the gate electrode. We can trace this behavior to the stacking-dependent response of the contact region to the capacitive electric field induced by the gate. The contact resistance is a central parameter and our observation establish an important design rule for devices based on 2D atomic crystals.
\end{abstract}

\maketitle

The recent advances in fabrication of heterostructures composed of stacks of atomically-thin layers bonded by van der Waals forces holds promise for ultra-thin electronics \cite{Novoselov2016,Dean2010,Kim2009,Wang2013}. A key-idea is to use the semi-metallic graphene (G) as an electrode material contacting a semi-conducting layer (2D-S) by a van der Waals bonded overlay region. A broad range of functional devices have been fabricated in this way including 2D field effect transitors (2D-FETs) \cite{Cui2015,Liu2015,Yu2014,Lee2015}, non-volatile memory cells \cite{Bertolazzi2013}, photoresponsive memory devices \cite{Roy2013}, and vertical tunneling FETs (V-TFETs) \cite{Georgiu2013}. 

The contact resistance due to the van der Waals gap and Schottky barrier between graphene and the 2D-S is, however, critical for device performance.
Opposed to conventional 3D metallic electrodes, the graphene work function ($W_\mathrm{G}$) and density of carriers ($n_\mathrm{G}$) are highly susceptible to external electric fields \cite{Yu2009,Yang2012}. As a result, in a gated 2D-S/G interface both the graphene electrodes as well as the semiconducting channel are affected by the gate field $\vec{E}_\mathrm{gate}$, so the contact characteristics ultimately depend on the collective response of the 2D stack to $\vec{E}_\mathrm{gate}$. 
An emblematic example is that of devices based on MoS$_\mathrm{2}$ with graphene contacts. Here several reports have observed a tunable contact resistance from rectifying to Ohmic by using increasingly positive gate voltages \cite{Bertolazzi2013,Cui2015,Liu2015,Yu2014,Lee2015}. In particular, linear I$_\mathrm{SD}$-V$_\mathrm{SD}$ characteristics, indicating the formation of an Ohmic contact between graphene and MoS$_\mathrm{2}$, have been observed in independent measurements at gate voltages V$_\mathrm{gate} \geq$ 80 V \cite{Liu2015,Cui2015}. Such behaviour has been associated with the modulation of the Schottky barrier at the contact induced by $\vec{E}_\mathrm{gate}$ \cite{Cui2015,Yu2014,Liu2015}. 
\begin{figure}
	\includegraphics[scale=0.125]{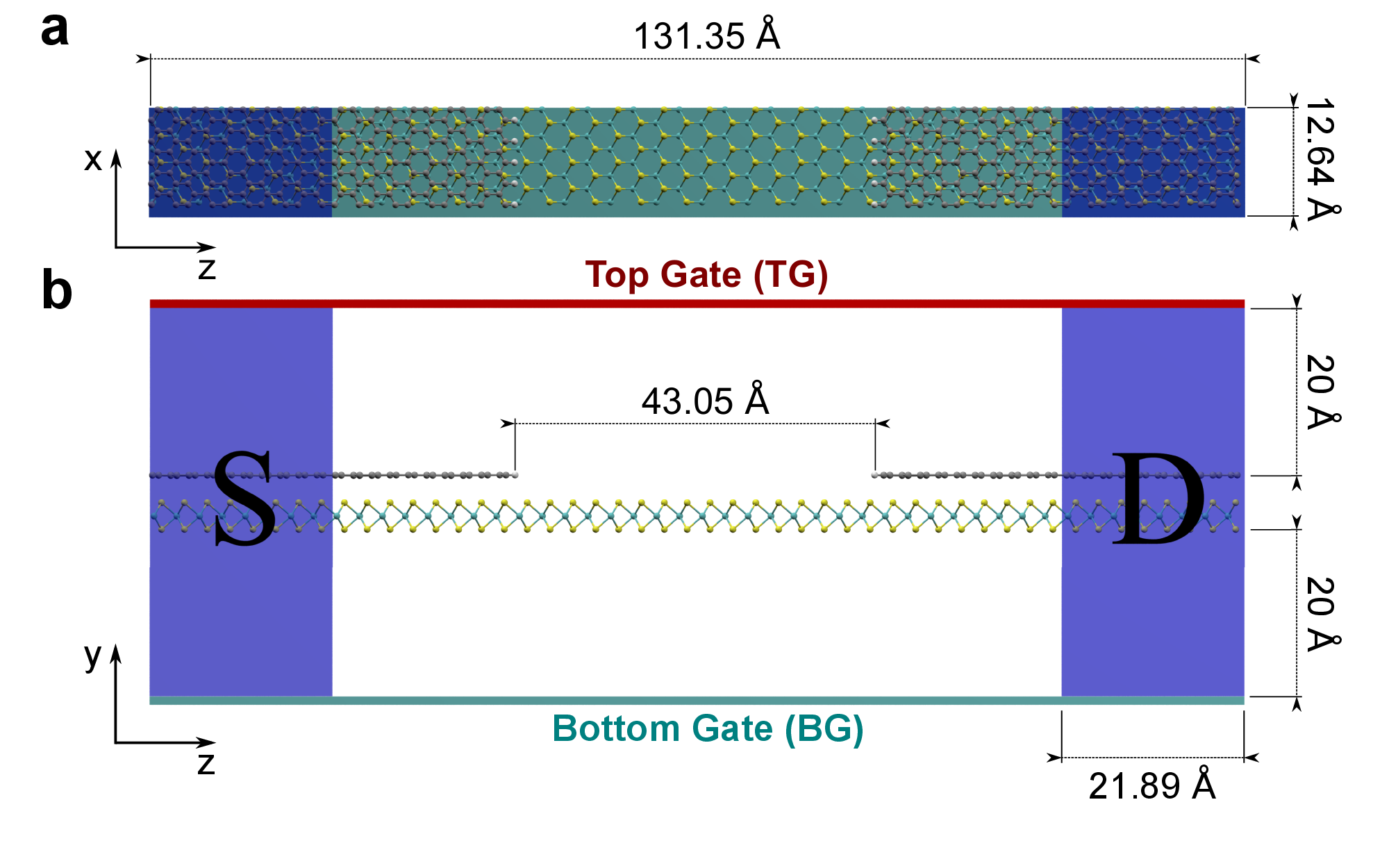}
	\caption{(Color online) Top (a) and side (b) view of the FET device setup used in the DFT-NEGF calculation. The semi-infinite G-\mosto\ overlay regions used as source (S) and drain (D) electrodes are highlighted in blue. The electrostatic planar top (TG) and bottom (BG) gates used to modulate the total charge in the structure are shown in red and blue, respectively.} 
	\label{fig:setup}
\end{figure}

The experimental results have been obtained with structurally different device architectures. In one setup the MoS$_\mathrm{2}$ is sandwiched in between the graphene and the gate electrodes (gate/MoS$_\mathrm{2}$/G setup) \cite{Cui2015,Yu2014,Lee2015}, whereas in others the graphene is positioned below the MoS$_\mathrm{2}$, in close proximity of the gate (gate/G/MoS$_\mathrm{2}$ setup) \cite{Bertolazzi2013,Liu2015}. While no apparent reason has been given for choosing one setup over the other, it has been argued that partial screening of $\vec{E}_\mathrm{gate}$ by MoS$_\mathrm{2}$ in gate/MoS$_\mathrm{2}$/G setups may lead to a larger contact barrier \cite{Liu2015}. The screening of $\vec{E}_\mathrm{gate}$ by MoS$_\mathrm{2}$ was also used to explain the saturation of $I_\mathrm{SD}$ with $V_\mathrm{gate}$ in vertical devices using a gate/MoS$_\mathrm{2}$/G setup \cite{Roy2013}. This calls for a  systematic characterization of the stacking-dependent response of the interface between graphene and 2D-SCs.  

Here we use first principles electron transport calculations based on density functional theory combined with non-equilibrium Green's functions (DFT-NEGF) to investigate the transconductance of a G-\mosto\ 2D-FET device at room temperature. We show how the stacking sequence matters: We find a significantly higher transconductance for a gate/\mosto /G device setup compared to a \mosto /G/gate one. The effect can be explained by considering the different screening characteristics of the two materials forming the contact. In the \mosto /G/gate setup, the influence of $\vec{E}_\mathrm{gate}$ on the \mosto\ work function ($W_\mathrm{MoS_2}$) is relatively small due to the good screening properties of doped graphene. On the other hand, in the gate/\mosto /G setup the \mosto\ is fully exposed to $\vec{E}_\mathrm{gate}$, leading to a stronger dependence of $W_\mathrm{MoS_2}$, and consequently, a stronger dependence of the Schottky barrier on gating.	

\begin{figure}
    \includegraphics[scale=0.7]{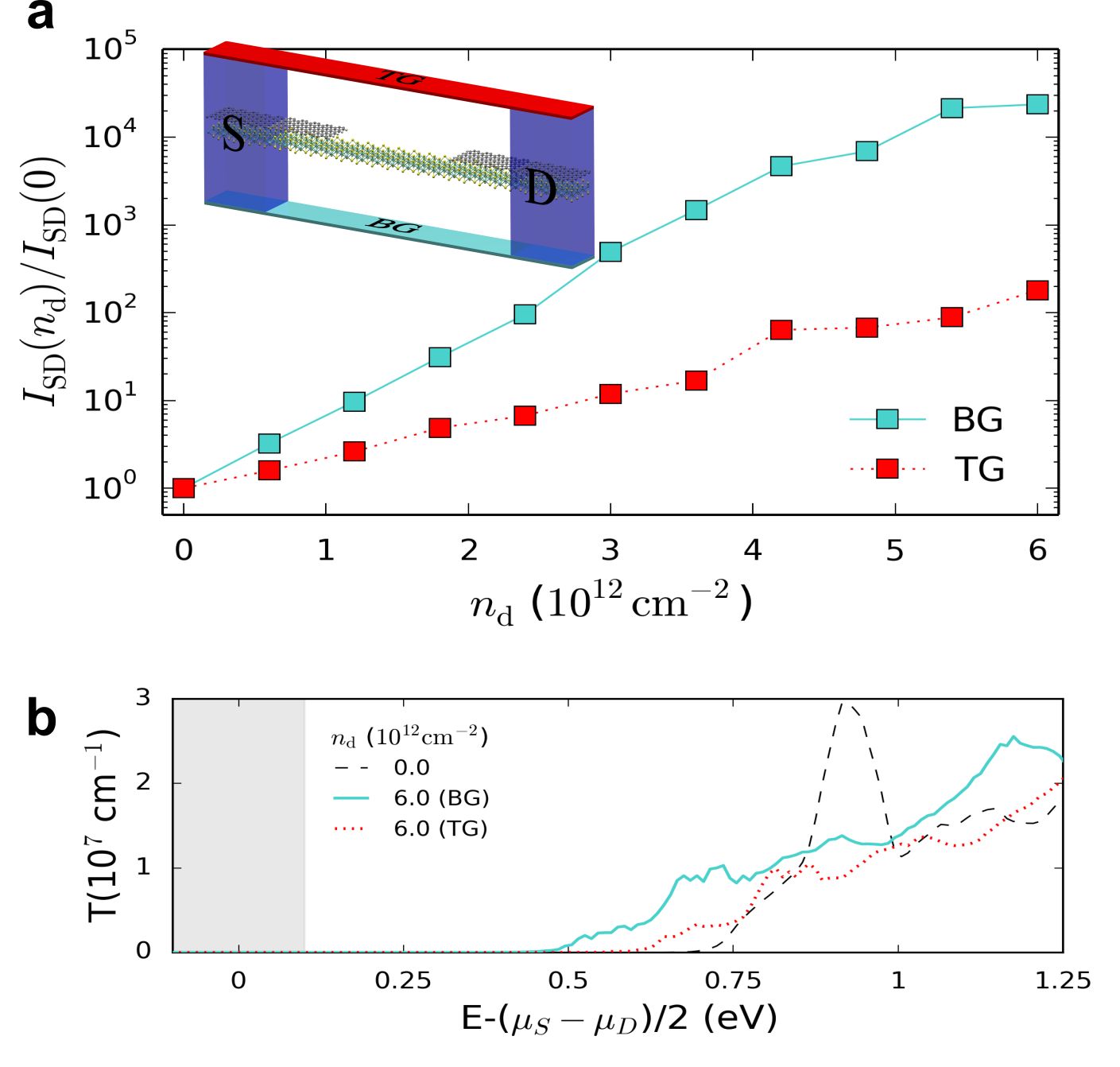}
    \caption{(Color online) (a) Source-drain current $I_\mathrm{SD}$ for the bottom-gated
        (turquoise, solid line) and top-gated (red, dotted
        line) device at source-drain bias $V_\mathrm{SD}=0.2\,\mathrm{V}$, as a
        function of the doping level $n_\mathrm{d}$. Inset: scheme of the device setup
        including the bottom gate (BG, turquoise) and top gate (TG, red). Carbon, Sulfur
        and Molybdenum atoms are shown in gray, yellow and cyan, respectively. The regions
        corresponding to the semi-infinite source (S) and drain (D) electrodes are shown
        as blue semi-transparent volumes. (b) Transmission spectra at $V_\mathrm{SD}$ =
        0.2 V and $n_\mathrm{d}$ = 6 $\times$ 10$^{12}$ cm$^{-2}$ for the TG (turquoise,
        solid line) and BG (red, dotted line) device setup. The black dashed line corresponds
        to the undoped case.}
  \label{fig:current}
\end{figure}

\begin{figure}
	\includegraphics[scale=0.4]{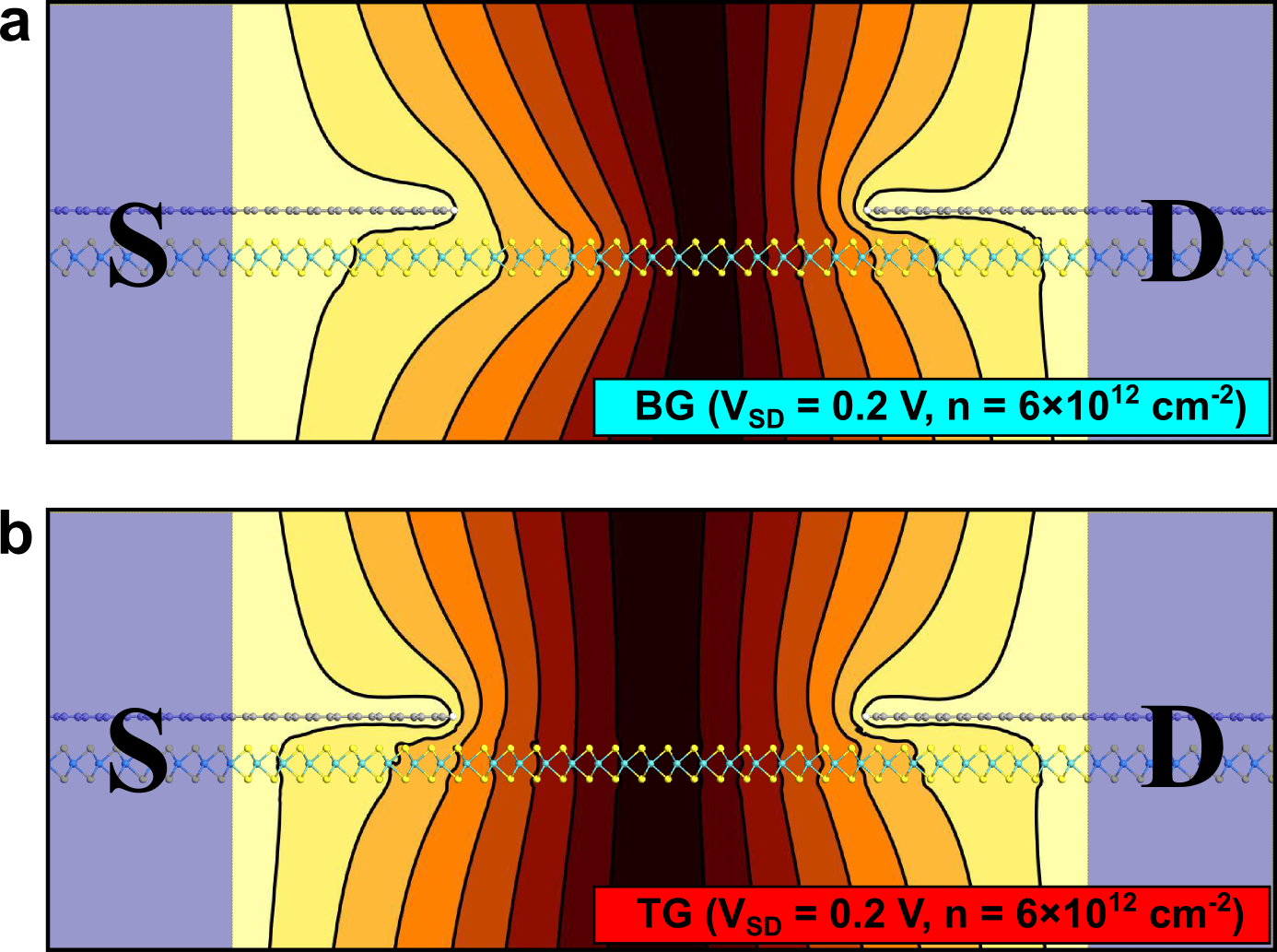}
	\caption{(Color online) 2D map of electrostatic potential drop at V$_\mathrm{SD}$= 0.2 V for BG (a) and TG (b). The data have been averaged along the X direction. For clarity, the map shows only the absolute value of the potential. Bright (dark) regions indicate regions of high (low) potential. The shaded blue regions indicate the S/D electrodes. Electrons move from S (left) to D (right). The contourlines are separated by 1.3$\times$10$^{-1}$ V.}
	\label{fig:voltagedrop}
\end{figure}

\emph{Setup and method ---} Our central device is shown in Fig.~\ref{fig:setup}. We
consider semi-infinite G-\mosto\ overlay regions as electrodes to a short \mosto\
channel. We have considered device setups in which electron transport occurs either along
the zig-zag (ZZ) or the armchair (AC) directions of graphene and MoS$_2$, with the
graphene terminations being AC or ZZ H-passivated graphene edges, respectively. Here we
show only results for the AC transport direction, as similar conclusions are reached also
for the ZZ case (see the Supplementary Material). The G-\mosto\ overlay structure is based on a 5$\times$5/4$\times$4 hexagonal
supercell in which the graphene is strained isotropically by 2.11\% and the \mosto\ is
kept at its equilibrium lattice constant (a$_\mathrm{MoS_2}$= 3.16 \AA). The interlayer
distance (C-Mo) is fixed to 4.9~{\AA}, as obtained from \textsc{Vasp} \cite{Kresse1996}
calculations using the DFT-D2/PBE method \cite{Perdew1996,Grimme2006} (see the Supplementary Material). The
charge in the device can be modulated by a gate electrode situated either below (bottom
gate, BG, gate/\mosto/G) or above (top gate, TG, \mosto/G/gate) covering the whole
structure.

The finite bias transconductance calculations have been performed using the \textsc{TranSiesta} \cite{Brandbyge2002,Papior2016b} DFT-NEGF code. Periodic boundary conditions have been applied in the direction transverse to the channel. The bulk heterostructure corresponding to the device electrodes has been calculated using the \textsc{Siesta} \cite{Soler2002} DFT code. We have used the PBE \cite{Perdew1996} functional, DZP (SZP) basis-set for graphene (\mosto), and 3/25 k-points in the transverse direction for electronic structure/transport calculations. The gating has been accounted for by introducing a planar region with a uniform charge distribution (charge gate, CG) \cite{Papior2016}. The associated Poisson equation used to obtain the Hartree term $V_\mathrm{H}$ in the DFT and DFT-NEGF Hamiltonians is

\begin{equation}
\nabla^2 V_\mathrm{H} (\mathbf{r}) =  -\frac{\rho_\mathrm{S} (\mathbf{r}) + \delta \rho_\mathrm{S} (\mathbf{r}) - \delta \rho_\mathrm{G} (\mathbf{r})}{\epsilon_0},
\end{equation}

where $\rho_\mathrm{S} (\mathbf{r})$, $\delta \rho_\mathrm{S} (\mathbf{r})$, $\delta \rho_\mathrm{G} (\mathbf{r})$ and $\epsilon_0$ are the electronic density of the non-gated system, the electronic density induced in the system by the gate, the corresponding counter-charge in the gate plane, and the vacuum permittivity. 

In order to access the CG method and the role of a dielectric screening inside the stack we performed
additional calculations using the \textsc{Atomistix ToolKit} package \cite{ATK}. In this case, the gate has been described by introducing a spatial region of constant $V_\mathrm{H}(\mathbf{r})$ (Hartree gate, HG). 
We modelled the encapsulating dielectric (diel.) layer by including a second spatial region in which the local Hartree potential $V_\mathrm{H}^\mathrm{diel.} (\mathbf{r})$  is determined from
\begin{equation}
\nabla^2 V_\mathrm{H}^\mathrm{diel.} (\mathbf{r}) = -\frac{\rho_\mathrm{S} (\mathbf{r})}{\kappa}.
\end{equation}
In the HG+diel. calculations, the dielectric permittivity has been set to $\kappa$ = 4$\epsilon_0$ to mimic encapsulation in an hBN stack. 

 \begin{figure}
  \includegraphics[scale=0.125]{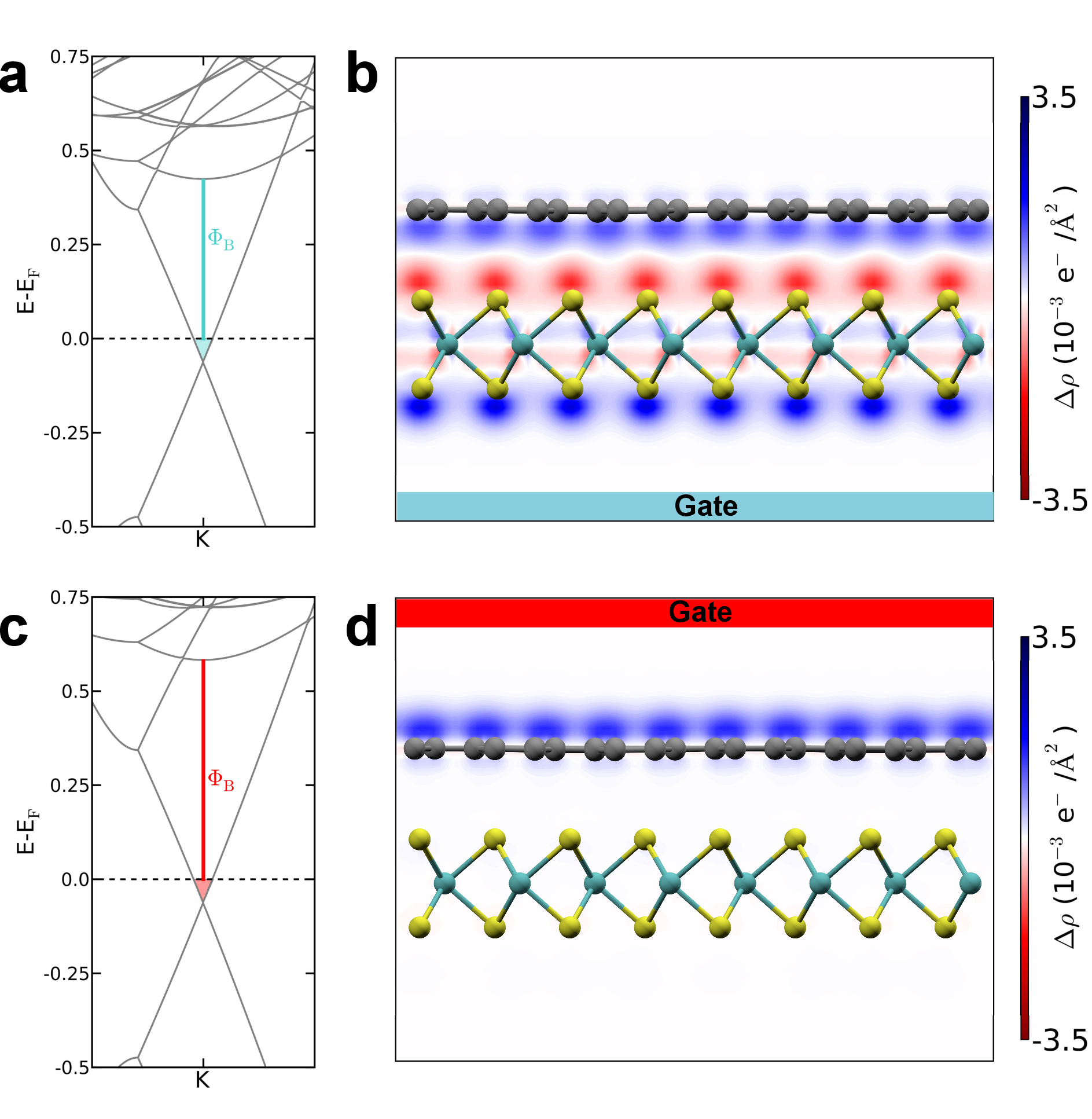}
  \caption{\small{(a) Electronic band structure around the K-point for the BG electrode configuration at $n_\mathrm{d}$ = 6.0 $\times$ 10$^{12}$ cm$^{-2}$. The energy on the Y-axis is scaled with respect to the Fermi energy E$_\mathrm{F}$. The Schottky barrier is indicated by the turquoise solid line. The shaded turquoise area indicates the portion of the graphene $\pi$* band below the Fermi energy. (b) Electronic density redistribution $\Delta\rho$ in the BG electrode configuration at $n$ = 6.0 $\times$ 10$^{12}$ cm$^{-2}$. $\Delta\rho$ has been integrated along the electrode short axis parallel to the graphene plane. Blue and red colors indicate electron accumulation and depletion, respectively.  (c,d) Same as (a,b), but for the TG electrode configuration.}}
  \label{fig:bands}
\end{figure}
\begin{figure}
  \includegraphics[scale=0.1]{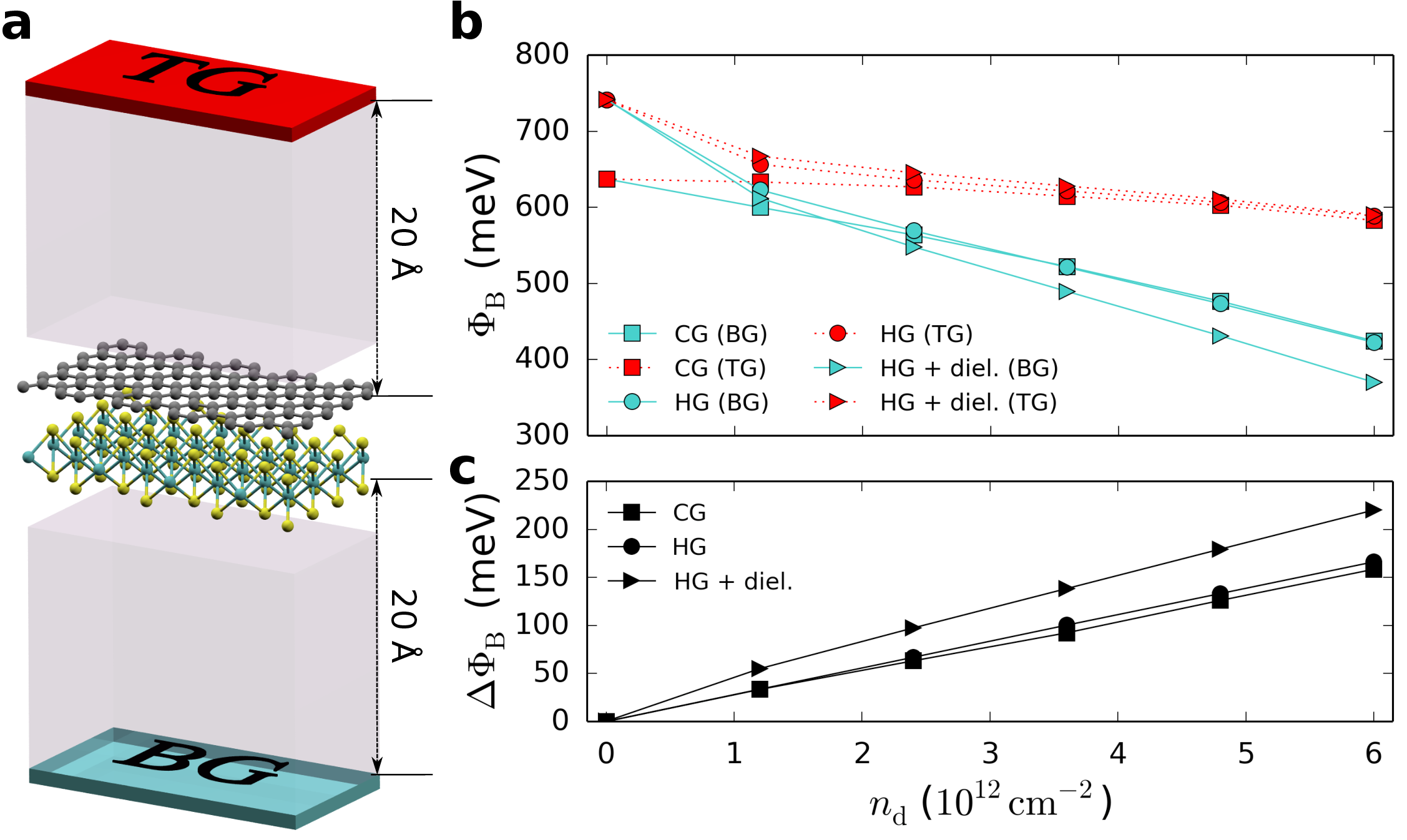}
  \caption{\small{(a) Geometry of the bulk electrode, including the bottom gate (BG, turquoise), the top gate (TG, red) and the dielectric region (purple). Carbon, sulfur and molybdenum atoms are shown in gray, yellow and cyan, respectively. (b) Schottky barrier height $\Phi_\mathrm{B}$ as a function of $n_\mathrm{d}$ for the bottom-gated (solid line, circles) and top-gated (dashed lines, squares) G-MoS$_\mathrm{2}$ bulk electrode calculated using SIESTA (squares), ATK (circles) and ATK including a dielectric region with $\kappa$ = 4$\epsilon_0$ to mimic encapsulation in hBN (triangles). (c) Difference in the Schottky barrier height $\Delta\Phi_\mathrm{B}$ as a function of $n_\mathrm{d}$.} }
  \label{fig:schottky}
  
\end{figure}
\emph{Results ---} In Fig.~\ref{fig:current}a we show the calculated $I_\mathrm{SD}$ for
an applied bias $V_\mathrm{SD}$ = 0.2 V and at room temperature. In agreement with
experimental observations \citep{Yu2014,Cui2015,Liu2015}, $I_\mathrm{SD}$ increases
considerably when an electron doping charge ($n_d$) is induced in the electrodes by the
gate.  However, the most striking feature is that the magnitude of $I_\mathrm{SD}$ differs
significantly depending on the active gate. For the TG device setup, where the graphene
overlaying electrode is closer to the gate, we observe markedly smaller currents where the
difference $\Delta I_\mathrm{SD}$ = $I_\mathrm{SD}$(BG) -- $I_\mathrm{SD}$(TG) increases
steadily with n$_\mathrm{d}$, reaching a factor of 100 at $n_\mathrm{d}$ = 6.0 $\times$
10$^{12}$ cm$^\mathrm{-2}$. Looking at the corresponding transmission functions
(Fig.~\ref{fig:current}b), we note that for the BG setup the on-set of transmission is
shifted to lower energies compared to the TG one, resulting in the larger current.  The
main scattering takes place at the junction between the electrode and \mosto\
channel. This can be seen from the voltage drop in Fig.~\ref{fig:voltagedrop} where the
drop is smaller at the S (positive) compared to the D (negative) electrode for the BG
setup while it is about symmetric for the TG one. This indicates that more negative charge
accumulates in the channel in the BG setup.

To shed light on the reason behind the dependence of $I_\mathrm{SD}$ on the stacking configuration, we have examined the electronic structure of the bulk heterostructure in the presence of TG/BG. Because of the weak van der Waals interaction between the graphene and \mosto, the electronic bands in the heterostructure can be regarded essentially as the superimposition of the electronic bands of the two individual 2D components. This allows us to evaluate the Schottky barrier at the contact, $\Phi_\mathrm{B}$, as the difference between the Fermi level $E_\mathrm{F}$ and the conduction band minimum (CBM) of \mosto\, see Fig. \ref{fig:bands}a,c. For the undoped G-MoS$_\mathrm{2}$ interface, $\Phi_\mathrm{B}$ = 640 meV, which is close to the value obtained with accurate $G_0W_0$ calculations \cite{Jin2015b,valueSB}. 

An intuitive picture of the dependence of $\Phi_\mathrm{B}$ on the stacking sequence can be drawn by examining the electronic density redistribution induced by gating in the MoS$_\mathrm{2}$/G heterostructure:
\begin{equation}
 \Delta \rho(\mathbf{r}) = \rho(\mathbf{r})[n_\mathrm{d} = 0\ \mathrm{cm^{-2}}] - \rho(\mathbf{r})[n_\mathrm{d} > 0\ \mathrm{cm^{-2}}]
\end{equation}
Upon gating, the additional carriers accumulate mainly in the graphene due to its metallic character. Thus, a capacitor develops in which the two plates are the graphene and the gate surface, respectively. In the BG configuration, the MoS$_\mathrm{2}$ lies within the capacitor, and its electronic states are strongly affected by the capacitive electric field $\vec{E}_\mathrm{gate}$, as shown in Fig.~\ref{fig:bands}b. This lowers the CBM of MoS$_\mathrm{2}$,\cite{Zibouche2014} leading to a strong decrease in $\Phi_\mathrm{B}$ compared to the ungated case. The scenario differs significantly in the TG configuration as clearly seen in  Fig.~\ref{fig:bands}d, as $\vec{E}_\mathrm{gate}$ remains confined within the gate and the graphene due to the good screening properties of the latter \cite{Stokbro2010}. As a consequence, $\vec{E}_\mathrm{gate}$ influences the MoS$_\mathrm{2}$ layer only weakly, leading to a much reduced dependence of  $\Phi_\mathrm{B}$ on $n_\mathrm{d}$  compared to the BG case. 

The Schottky barriers, $\Phi_\mathrm{B}$,  calculated from the bandstructures at different $n_\mathrm{d}$  are shown in Fig.~\ref{fig:schottky}b. 
Gating the heterostructure leads to an overall decrease of the Schottky barrier, $\Phi_\mathrm{B}$. This decrease is considerably faster for the BG configuration than for the TG one. Indeed, at $n_\mathrm{d}$ = 6.0 $\times$ 10$^{12}$ cm$^\mathrm{-2}$, $\Phi_\mathrm{B}$(BG) = 420 meV, whereas $\Phi_\mathrm{B}$(TG) = 580 meV. This trend is perfectly consistent with that of $I_\mathrm{SD}$ {\em vs.} $n_\mathrm{d}$, and with the shift in on-set of electron transmission shown in Fig.~\ref{fig:current}b. We may estimate the ratio of the thermionic currents at $n_\mathrm{d}$ = 6.0 $\times$ 10$^{12}$ cm$^\mathrm{-2}$ to be $e^{\Delta\Phi_B/k_B T}\sim 600$, in reasonable agreement with  Fig.~\ref{fig:current}a. 

In addition to $\Phi_\mathrm{B}$, the tunnelling current in the gated G-MoS$_\mathrm{2}$ contact depends also on the number of graphene carriers ($n\mathrm{_G}$) available for injection into the MoS$_\mathrm{2}$. The latter can be extracted from the relation $E_\mathrm{D}$ = $\hbar \upsilon_F \sqrt{\pi n\mathrm{_G}}$, where $E_\mathrm{D}$ is the energy of the Dirac point with respect to $E_\mathrm{F}$  and $\upsilon_F$ = 10$^6$ m s$^{-1}$. Within the range of doping considered, $n_\mathrm{G}$ (BG) $\approx$ $n_\mathrm{G}$(TG), and even at $n_\mathrm{d}$ = 6.0 $\times$ 10$^{12}$ cm$^\mathrm{-2}$,  the Dirac point is shifted by a similar amount in the BG and TG setups (see Fig.~\ref{fig:bands}a,c). This indicates that the response of the FET device is dominated by the different modulation of $\Phi_\mathrm{B}$ in the two device setups considered. 

In Fig.~\ref{fig:schottky} we also compare the $\Phi_\mathrm{B}$ {\em vs.} $n_\mathrm{d}$ data obtained with the CG method with those obtained with the HG and HG+diel. methods. The three methods yield essentially the same result for the chosen gate distance of 20 {\AA}, the qualitative trend of $\Phi_\mathrm{B}$ with $n_\mathrm{d}$ being very similar with only minor differences in the actual values of $\Phi_\mathrm{B}$ due to the slightly different computational setups employed.

\emph{Conclusions ---} We have performed first principle DFT-NEGF calculations on a short \mosto\ device contacted by graphene showing how the contact resistance between overlayed graphene electrodes and the 2D semiconductor is very sensitive to the position of the gate. This points out a novel design rule for future electronic devices based on stacked heterostructures.

We thank Dr. F. Pizzocchero, L. Gammelgaard, B. S. Jessen, K. Stokbro and Profs. P. B{\o}ggild, K. S. Thygesen for discussions. The Center for Nanostructured Graphene (CNG) is sponsored by the
Danish Research Foundation, Project DNRF103. DS acknowledge support from the HC\O\  DTU-COFUND program.
 
\bibliography{biblio}

\end{document}